\documentclass[10pt,prl,twocolumn,aps,showpacs,superscriptaddress]{revtex4-1}
\usepackage{epsfig,amsmath,amssymb,graphicx,color,calc,epstopdf}

\def\tr{{\rm tr}}

\def\to{\rightarrow}

\newcommand{\beq}{\begin{equation}}
\newcommand{\eeq}{\end{equation}}
\newcommand{\ba}{\begin{align}}
\newcommand{\ea}{\end{align}}

\begin{document}
\title{The simplest model of jamming}
\author{Silvio Franz}
\affiliation{Laboratoire de Physique Th\'eorique et Mod\`eles Statistiques, Universit\'e Paris-Sud 11 and CNRS UMR8626, Batiment 100, 91405 Orsay, France}

\author{Giorgio Parisi}
\affiliation{Dipartimento di Fisica,
Sapienza Universit\`a di Roma,
INFN, Sezione di Roma I, IPFC -- CNR,
Piazzale Aldo Moro 2, I-00185 Roma, Italy
}

\begin{abstract}
  We study a well known machine learning model -the perceptron- as a
  simple model of jamming of hard objects.  We exhibit two regimes: 1)
  a convex optimisation regime where jamming is hypostatic and
  non-critical. 2) a non convex optimisation regime where jamming is
  isostatic and critical.  We characterise the critical jamming phase
  through exponents describing the distributions law of forces
  and gaps. Surprisingly we find that these exponents coincide with
  the corresponding ones recently computed in high dimensional hard
  spheres.  In addition, modifying the perceptron to a random linear
  programming problem, we show that isostaticity is not a sufficient
  condition for singular force and gap distributions. For that,
  fragmentation of the space of solutions (replica symmetry breaking)
  appears to be a crucial ingredient. We hypothesise universality for
  a large class of non-convex constrained satisfaction problems with continuous
  variables.
\end{abstract}


\maketitle


\paragraph{Introduction--} 
Jamming of hard objects is a general phenomenon that has attracted lot
of interest, both experimental and theoretical, (see
\cite{NLSW10,TS10} for recent reviews).  Jamming of hard spheres,
where the only interaction is excluded volume, has been widely
studied: the jamming point is reached when{, both under equilibrium or
off-equilibrium conditions,} the size of the cages where
the particles can move shrinks to zero. In this limit the system is 
critical: the network of contacts is isostatic \cite{M98,TV99}, 
 and many quantities have an anomalous power behaviour with
non-trivial critical exponents \cite{OLLN02,OSLN03,W12}. 
{ While local
excitations give rise to exponents that may depend on the spatial
dimension \cite{LDW12,LDW13}, the contributions of long range
excitations appear to be super-universal: numerical simulations show
that upon removing the contributions of local excitation, the critical
exponents have a very weak dependence on the space dimension
in wide range of dimensions \cite{fractal,CCPZ14}. Moreover, these
exponents seem to be independent from the protocol used to generate
jammed configurations.  } In the infinite
dimensional limit one expects some simplifications to be
present and one can study analytically jamming at equilibrium. 
Indeed (if we disregard crystallisation) the thermodynamics
of a gas of thermal hard spheres can be solved when the dimension goes
to infinity \cite{KPZ12,KPUZ13,CKPUZ14,CCPZ12}. One finds a rather
unexpected scenario:

 At low pressure (low density) we stay in the liquid phase.
Increasing the density at a pressure $P_g$ we enter into the glass phase where the
spheres are confined into small volume cages.

 Further increasing the density, at a high pressure $P_G$ we
  enter into a different glass phase where the cages split into smaller
  cages when increasing the pressure. This process goes on up to
  infinite pressure.

In the infinite pressure limit one finds analytically the power
  laws that have been discovered 
in  \cite{WSNW05,W12,IBB13}: the exponents can
  be computed and they are compatible (within the
  numerical bounds) with the exponents found in dimensions from 2
  upwards. 

  From an abstract viewpoint the problem of packing spheres in space
  can be viewed as a constraint satisfaction problem where one aims to
  maximize the sphere's packing fraction subject to the hard core
  impenetrability constraints. Statistical physics has been
  instrumental in the analysis of random constraint satisfaction
  problems, where the constraints on the variables are chosen at
  random from an ensemble: by changing the number or the nature of the
  constraints we go from a satisfiable (SAT) phase (where there is at
  least one configuration that satisfies all the constraints) to an
  unsatisfiable (UNSAT) phase (where it is impossible to satisfy all
  the constraints at the same time). Usually this SAT-UNSAT transition
  becomes sharp in the thermodynamic limit: with the exclusion of the
  transition point, the probability of a random problem to be
  satisfiable becomes 0 or 1 in the thermodynamic limit where the
  number of the variable of the systems ($N$) and the number of
  constraints ($M$) goes to infinity at fixed ratio $\alpha=M/N$.
  Random constraint satisfaction problems have been widely studied by
  physicists in the case of discrete variables: the most celebrated
  case is the random K-SAT problem \cite{MPZ02}, where the physicist's
  solution has been recently proved for sufficiently large $K$
  \cite{DSN14}. Polydisperse hard spheres \footnote{In the limit of
    zero polydispersitivity the randomness disappears, but one can
    still use the standard tools of statistical mechanics as replica
    and cavity methods in this limit.} are a particular case of random
  constraint satisfaction problem, but they differ from the mostly
  studied cases by the nature of the variables which are continuous
  rather than discrete. The continuous nature of the variables adds a
  new dimension to the the problem: the SAT-UNSAT transition coincides
  with an equilibrium jamming transition where the volume of the space
  of the satisfying assignments of variables shrinks to zero and the
  system can become critical.

The infinite dimensional limit of hard spheres is the first example of
non-trivial continuous constraint satisfaction problems that has been analytically
solved. One may wonder how much general is the critical picture of 
the jamming transition, and if criticality there is, if a unique or
several universality classes are possible. 
The aim of this letter is to present a simple model of jamming, that can be
solved analytically and it has the same jamming exponents of hard
spheres. The model, the spherical perceptron, is very well known in
machine learning and neural network theory, where it is used as
linear signal classifier \cite{R58}. 

\paragraph{A toy sphere packing problem}
We start from an extreme schematisation of sphere jamming problem in
which we substitute the interaction between spheres with a random
background and consider a single particle that should not overlap with
some spherical obstacles placed in random positions in space. Both the obstacle 
and the single particle live on $S_N$, the $N$ dimensional sphere of radius $\sqrt{N}$.  Let us
consider $M=\alpha N$ point obstacles  $\xi_i^\mu$ ($i=1,...,N$
$\mu=1,...,M$) in fixed random positions
on the $N$
dimensional sphere, and a particle at $x_i$ constrained to be at a
distance greater than $\sigma$ from the obstacles: $|\xi^\mu-{\bf
  x}|>\sigma$. 
As we shall see this problem is isomorphous to the perceptron.

\paragraph{The perceptron --}
Also in this case the configuration space is the sphere $S_N$ of normalised
$N$-dimensional vectors ${\bf x}$ such that $\sum_{i=1,N} x_i^2=N$.
We impose the following $M=\alpha N$ constraints:
we have $M$ $N$-dimensional random vectors ${\bf \xi}^\mu$ (with the same
normalisation)  and we require that
\begin{equation}
r_\mu\equiv \frac{1}{\sqrt{N}}\sum_i \xi^\mu_i x_i - \kappa>0 \ \ \ 
\forall \mu =1,...,M \,.
\label{uno}
\end{equation}
The quantities $r_\mu$ will be called gaps in the following.  In a
packing perspective we see that the problem of the toy spheres
coincides with the perceptron for $\sigma^2=2 N+2 \kappa$.
{ The analogue of packing fraction maximisation in the problem of hard
  spheres
  is here maximisation of $\kappa$ for fixed $\alpha$. } While in
  machine learning the interest is generally restricted to the
  positive values of $\kappa$ (see \cite{S13} for an exception), for
  the jamming problem negative values are equally legitimate and, we
  will see, more interesting. In the $\alpha$-$\kappa$ plane there is
  a SAT region where the previous inequalities have at least one
  solution (with probability one) and an UNSAT region where there is
  no solution.  The line of jamming points that separates the two
  regions has the shape shown in fig. 1.  The value of
  $\alpha_c(\kappa)$ on this line is usually called the maximum
  perceptron capacity in machine learning. The perceptron problem has
  been studied in the past with statistical physics approaches
  \cite{G88,GD88} and the results of Gardner and Derrida for
  $\kappa\ge 0$ are well know: defining $D_{\sigma^2} y$ the Gaussian
  measure with zero average and variance $\sigma^2$ ($Dy\equiv D_1y$),
  and the error function $H(x)=\int_x^\infty Dy$, we have
\begin{eqnarray}
\label{alpha}
\alpha_c(\kappa)=\left( \int_{-\kappa}^\infty Dh (h-\kappa)^2 \right)^{-1}
\end{eqnarray}
The distribution of the gaps $g(r)$ at jamming, that we normalize to the ratio of the
number of constraints over variables $\alpha$, is given by 
\begin{eqnarray}
  \label{eq:15}
  g(r)=\alpha (1-H(\kappa))\delta(r)+\frac{\alpha}{\sqrt{2 \pi}} e^{-(r+\kappa)^2/2}\theta(r).
\end{eqnarray}
According to equations (\ref{alpha},\ref{eq:15}), for $\kappa>0$ the system is
'hypostatic': the weight of the delta function contribution in
(\ref{eq:15}) gives precisely the fraction of binding constraints in
the system, for which (\ref{uno}) is verified as an equality. This is 
a decreasing function of $\kappa$, smaller than one for $\kappa>0$ and
equal to one
exactly at $\kappa=0$. Notice that the $r>0$ part of the distribution
is regular and has a finite limit for $r\to 0$. These results contrast with
salient features of jamming in hard spheres: the property
of isostaticity and the presence of a power law singularity in the gaps
distribution at small $r$ \cite{OLLN02,OSLN03}.

However these formulae are valid only for positive or zero $\kappa$
($\alpha(0)=2$).  In this situation, for any $\alpha<\alpha_c(\kappa)$
the space of allowed configurations is convex on the sphere (it is the
intersection of convex domains) and the situation is well under
mathematical control \cite{T11,S13}.  The case of negative $\kappa$ is much harder,
each constraint defines a non-convex allowed domain and the space of
solutions can fragment in disconnected regions. In the
statistical approach that we are going to describe below, this
phenomenon corresponds to replica symmetry breaking, while the previous
formulae have been derived in a replica symmetric assumption. We will show
that as soon as $\kappa<0$ replica symmetry is broken and critical
universality at the jamming point emerges, the system is isostatic
(i.e. the number of contacts is equal to the dimension of the space,
i.e. $N$), $g(r)$ displays a power law singularity $g(r)\sim
r^{-\gamma}$ at small $r$, which is in turn associated to a pseudo-gap
in the force distributions $P(F)\sim F^\theta$ at small $F$.
{  With an argument analogous to the one used in hard spheres \cite{W12},
one can show (see SM) that the exponents verify the stability bound
$\gamma\ge 1/(2+\theta)$. The arguments of \cite{MW14} can be used to
argue that stability should be marginal and this inequality
saturated.  The values we find $\gamma=0.41269$ and
$\theta=0.42311$ indeed saturate to numerical precision the bound
and they coincide with the ones found in high dimensional hard
spheres. }

In order to study the model it is convenient to introduce an
Hamiltonian ${\cal H}(x)$ that is non zero only if all the constraints are
violated. A choice analogous to the soft sphere Hamiltonian is
\begin{equation}
\label{H}
{\cal H}(x)=\frac 1 2 \sum_{\mu=1}^M r_\mu^2\theta(r_\mu)\,,
\end{equation}
where $\theta$ is the standard Heaviside function. 
In the following we will concentrate mainly 
on analytic computation of
the Gardner volume \cite{G88} of the satisfying assignments
\begin{eqnarray}
  \label{eq:19}
  {\cal V}(\alpha,\kappa)\equiv e^{NS(\alpha,\kappa)} = 
\int_{S_N} d{\bf x}\prod_{\mu=1}^{\alpha N} \theta(r_\mu)
\end{eqnarray}
{  The equilibrium jamming transition line is the locus of points where this volume
shrinks to zero, and we will approach it from the SAT phase.
To test our theoretical findings, and show that also in the
  perceptron the critical properties of jamming are independent of the
  preparation of the jammed configurations, we will also present
  numerical simulations, where we generate non-equilibrium jammed
  configurations through local minimisation of ${\cal H}$. }
  {  It is important to distinguish equilibrium jamming transition
  from off-equilibrium jammed configurations generated with heuristic
  minimisation algorithms. These last can be defined as isolated
  points of minimum of ${\cal H}$ with ${\cal H}=0$. } 

\paragraph{The Gardner Volume}
The quenched average of the entropy $S(\alpha,\kappa)$ over the random 
vectors ${\bf \xi}^\mu$ can be performed with replicas. Alternatively
the more transparent but cumbersome cavity method could be used \cite{M89}.
Following standard computations, see e.g. \cite{MPV}, one finds
\cite{G88}  that
the entropy can be expressed as a saddle point over the overlap matrix
between solutions in different replicas 
$Q_{a,b}= N^{-1}  \sum_i \langle x^a_i x^b_i\rangle$
where $a,b=1,...,n$ with $n\to 0$ at the
end \cite{GR97}:
\begin{eqnarray}
  n S[Q]=1/2 \tr \;\log \; Q + \\ \alpha \log \left.\left( e^{ \frac 1 2 \sum_{ab} Q_{ab} \frac{\partial ^2}{\partial h_a\partial h_b} }\prod_a \theta(h_a-\kappa)\right)\right|_{h_a=0}\nonumber
\label{srep}
\end{eqnarray}
Assuming that the replica symmetry is broken in the usual hierarchical
ultrametric way \cite{MPV}, one gets an explicit form of the entropy that we
report for reader convenience.  First of all, using the parameterization the
matrix $Q$ in terms the function $x(q)$ which varies in an interval
$[q_0,q_1]$ \cite{MPV}, we have
\begin{eqnarray}
  \label{eq:1.5}
&& \frac{1}{n} \tr \;\log \; Q =\log(1-q_1)+\frac{q_0}{\lambda(q_0)}+\int_{q_0}^{q_1} dq\; \frac {1}{\lambda(q)}\\
&&\lambda(q)=1- q_1 +\int_q^{q_1}dq'\; x(q').
\end{eqnarray}
Secondly, the remaining term in the entropy, can be written as $-n \alpha\int
D_{q_0}(h-\kappa) f(q_0,h)$, where, indicating with dots
$q$-derivatives and with primes $h$-derivatives, the function $f(q,h)$
verifies the partial differential equation \cite{P80}:
\begin{eqnarray}
  \label{eq:2}
\dot{f}=-\frac{1}{2}(f''+xf'^2)
\end{eqnarray}
with boundary condition
\begin{eqnarray}
\label{bcf}
f(q_1,h)=
-\log H\left( \frac{\kappa-h}{\sqrt{1-q_1}}\right).
\end{eqnarray} 
As usual, in order to write the variational equations for $x(q)$ it is useful to 
define the distribution of the local gaps at level $q$: $P(q,h)$ which
verifies \cite{SD84,GR00} 
\begin{eqnarray}
  \label{eq:3}
\hspace{-2 mm}
\dot{P}=\frac{1}{2}(P''-2x(mP)')\ ;  \ \ \
  P(q_0,h)=D_{q_0}(h)/dh \,,
\end{eqnarray}
where we introduced $m(q,h)=f'(q,h)$
that verifies
\begin{eqnarray}
\label{eq:8}
\dot{m}=-\frac{1}{2}(m''+2x m m').
\end{eqnarray}
The variational equations with respect to $x(q)$ read
\begin{eqnarray}
  \label{eq:4}
\frac 1 2 \left(\frac{q_0}{\lambda(q_0)^2}+\int_{q_0}^qdq'\;
  \frac{1}{\lambda(q)^2} \right)-\frac \alpha 2 \int dh\; P m^2=0.
\end{eqnarray}
The RS solution is recovered from the above formulation in the limit
$q_1=q_0$. If there is a continuous part in $x(q)$ it is useful
to consider the derivatives of the (\ref{eq:4}) w.r.t. $q$  
 \begin{eqnarray}
  \label{eq:5}
  \hspace{-2 mm} 
 \frac{1}{2 \lambda(q)^2}-\frac{\alpha}{2}\int dh\; P m'^2=0 \,.
\end{eqnarray}
which, as it is well known, signals that continuous
RSB solutions are marginally stable with a divergent spin glass susceptibility.  In any
discontinuous solution, stability requires positivity of the l.h.s. of
(\ref{eq:5}).
For each value of $\kappa$ at sufficiently low values of $\alpha$ the
system is in the replica symmetric ``liquid'' phase: the space of SAT
assignments is simply connected and one can go with continuity from
one solution to the others. At higher values of $\alpha$ replica
symmetry breaks down and the space of solution becomes
disconnected. The line of transition for $\kappa<0$, along with the
jamming line estimated from the RS solution, are presented in
fig. 1. The RSB transition occurs in the SAT phase for $\kappa<0$, and as
announced, the jamming line lies in the glassy RSB phase.  Generically,
the RSB solution space fragmentation can occur either through a second
order transition to a continuous RSB phase or through a
discontinuous Random First Order Transition \cite{MPV}.  Close to
$\kappa=0$, and down to $\kappa=\kappa_{1RSB}\approx-2.05$ one finds
RSB to a continuous solution occurring via a de Almeida-Thouless
\cite{AT78} instability of the RS solution. Below the value
$\kappa_{1RSB}$ one finds a transition to a discontinuous ``one step''
solution. It is important to remark however, that upon increasing
$\kappa$ at fixed $\alpha$ so to approach jamming, a second transition
to a continuous solution, the so called Gardner transition should be
expected, so that for all $\kappa<0$ jamming is described by a
continuous solution.

\paragraph{Jamming} The jamming transition is the point where the space of solutions
 shrinks to a point, the entropy goes to $-\infty$, and the solution's
 self-overlap $q_1\to 1$.  
In this conditions, the boundary condition (\ref{bcf})
 for $f$ in $q_1$ reduces to
 \begin{eqnarray}
   \label{eq:17}
 f(q_1,h) \approx \frac{-h^2\theta(-h)}{2(1-q_1)} \;\;\; q_1\to 1. 
 \end{eqnarray}
 As for jamming in spheres or for the low temperature limit of the SK
 model, we can expect a scaling regime to emerge where all the
 relevant values of $q$ are close to unity.
\begin{figure}
\epsfxsize=250pt 
\epsffile{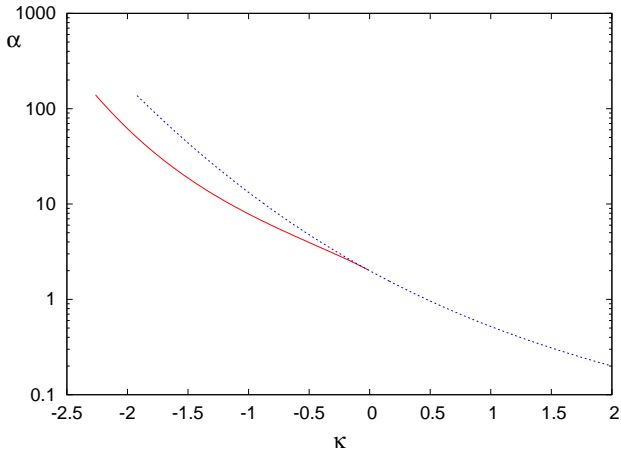}
\caption{Phase diagram of the model. The de Almeida-Thouless
  (AT) line of instability of the RS solution in the $\alpha-\kappa$
  plane (red line), together with the jamming line (blue) as estimated
  from the RS solution.  The jamming line is exact for $\kappa>0$ while
  corrections are to be expected for $\kappa<0$, where the RS solution
  just provides an upper bound to the true value \cite{S13}.}
\label{fig1}
\end{figure}
The existence of the scaling regime is intimately related to the
behaviour of the distributions of the small gaps and forces $g(r)\sim r^{-\gamma}$ and $P(F)\sim
F^{\theta}$. As described in more detail in
the SM, the scaling regime can be studies solving first the equations
for $P$ and $m$ for large positive and negative values of $h$, and
then matching both regimes via scaling functions
$\Delta^{-a/k}p_1(h\Delta^{-1/2})$ and
$\frac{\Delta^{1/2}}{\lambda(\Delta)} m_1(h\Delta^{-1/2})$ where
$\Delta=1-q$. The exponents $a$, and $k$ are determined from the
explicit solution to the equations. It turns out that $p_1$ satisfy
the same equation and boundary conditions of the analogous function
that appears in the hard sphere problem, and that therefore, scaling
function and exponents are the same to this case. The difference
between the two cases comes in the distributions of gaps and forces at
large values that are non-universal. 

The behaviours of $p_1(u)$ at large positive and negative argument are
related respectively to the small gap and the small force distributions 
and behaves as 
\begin{eqnarray}
p_1(u) \sim 
\left\{
\begin{array}{ll}
u^{-\gamma}& u\to \infty^+\\
|u|^{\theta}& u\to \infty^-
\end{array}
\right.
\end{eqnarray}
with $\gamma = 2 a/k=0.41269$, $\theta=\frac{k-1-a}{1-k/2}=0.42311$. 
The relation $\gamma=1/(2+\theta)$ is verified within numerical
errors and isostaticity holds.

\begin{figure}
\epsfxsize=250pt 
\epsffile{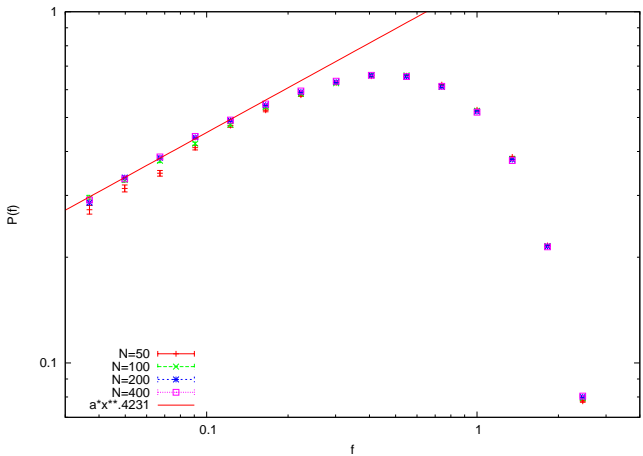}
\caption{The distribution of forces at jamming for $\alpha=4$ and
  $N=50,100,200,400$ in a log-log scale. The straight line is a one
  parameter fit of the kind $P(f)=a f^\theta$ with the predicted value
  of $\theta=0.42311$. }
\label{fig2}
\end{figure}

\paragraph{A simple variant: random linear programming} 
Our results are quite surprising, the jamming transitions in the
perceptron and in hard spheres, which are rather different problems
are in the same universality class.  Both in the high dimensional
spheres and in the present case, {  isostaticity as well as} the singular
behaviour of the gap and force distribution at small argument appear as
non-trivial consequences of the full-RSB solution to the problem. {  It
is therefore natural to ask if RSB is a
necessary ingredient, if isostaticity and power laws are always
associated or there are models where isostaticity hold while
the system remains replica symmetric, and if in this case the singular
behaviour in $g$ and $P$ is found.}  To address these questions we
slightly modify the problem (\ref{uno}). On the one hand we relax the
spherical constraint on ${\bf x}$, which then becomes a vector of $N$
real variables, on the other we modify the constraints to:
\begin{equation}
\label{rlp}
\frac{1}{\sqrt{N}}\sum_i \xi^\mu_i x_i+s^\mu - \kappa>0 \ \ \ 
\forall \mu =i,M \,.
\end{equation}
where, as before the $\xi^\mu$ are vectors on the sphere and we
introduced new constants $s^\mu$ that we take as Gaussian numbers of
zero average and variance $\sigma$. The maximisation of $\kappa$, and
the simultaneous determination of the maximiser vector ${\bf x}$ is a
linear programming problem which is convex for all values of
$\alpha$. For $\alpha<2$ and all $\kappa$ the constraints (\ref{rlp})
define an open region of space and no finite maximum for $\kappa$
exist.  We concentrate then on $\alpha\ge 2$ where the constraints
define a closed region of space and a maximum $\kappa$ exists. The
analysis of the Gardner volume for this model is very similar to the
one of the perceptron (see SM) with the important difference that here
replica symmetry always holds.  Remarkably, the RS solution is
marginally stable and isostaticity holds. { It seems a
  general property that isostaticity is associated to marginal
  stability of the replica solution, but as this example shows,  not necessarily to replica
  symmetry breaking.} However, differently from the perceptron for
negative $\kappa$, the distribution $g(r)$ is regular and finite at
$r=0$. We conclude that singular behaviour of the distributions and
fragmentation of the space of allowed configurations are intertwined
phenomena.
\paragraph{Simulations --}
In order to check the soundness of these predictions we have performed
the following numerical experiment. We have started from a random
configuration at $\alpha=3$ and we have found a minimum of the
Hamiltonian (\ref{H}) in the UNSAT region $\kappa>\kappa_c(\alpha)$, and we have
then approached the jamming point decreasing the value of $\kappa$
 until the point where
the energy is about $10^{-12}$. We obtain in this way a jammed
configuration with $\kappa=\kappa^*\approx -0.431$ (the value we
obtain slightly depends on the search procedure). Notice we did not try to equilibrate
the system: the jammed configuration that we reach can be expected to
be different from the one computed analytically studying the Gardner
volume and $\kappa^*<\kappa_c(3)$.  We can however study the
probability distribution of the gaps and of the forces in the
configurations found in this way. We have done this analysis for moderate
values of $N$ (in the range $50-400$), and found that for both
quantities the distributions at small values are compatible with the
thermodynamical predictions. { This is same qualitative coincidence of
exponents in equilibrium and off-equilibrium that is observed in
the field distribution of SK model at zero temperature \cite{P03,DMW10} 
and in hard spheres.}
The data of best quality are these for
the force distribution and are presented in figure (\ref{fig2}). 
We also simulated in a similar way the random linear programming
model, where thanks to convexity the limit of thermodynamic jamming
can be readily reached, and we find force and gap distributions in
perfect agreement with the RS predictions with no singularity at small
argument.

\paragraph{Conclusions --}

The most important results of this letter could be summarised as
follow: jamming in the perceptron is hypostatic and stable in the
region $\kappa>0$ where it defines a convex optimisation problem,
conversely in the non convex region $\kappa<0$ it is isostatic and
marginally stable. Remarkably jamming is in the same universality
class as hard spheres. The salient feature of jamming of a singular
behaviour of the gap and force distribution appears to be critically
related to fragmentation of the space of configurations -continuous
RSB- and the existence of a scaling solution with $q\approx 1$ on
approaching the jamming line. The random linear programming problem
shows that isostaticity is not {  necessarily associated to
}  that
behaviour. One finds there a jamming line which is isostatic and
marginally stable  which 
remains however Replica Symmetric, and the gap and force distributions
are regular at small argument. On the bases of these finding we
formulate the conjecture that universal jamming behaviour occurs in a
large class of non convex continuous random CSP with inequality
constraints. This could be tested e.g. generalising the perceptron
problem to correlated positions of the obstacles $\xi^\mu$, or
restricting the range of allowed values of the $r^\mu$ to finite
intervals or also, considering diluted perceptrons where for each
given $\mu$ only a finite number of $\xi^\mu_i$ are allowed to be
non-zero etc. It could also be interesting to see if the universality
extends to systems with constraints of a different nature, e.g.  where
the number of non-zero $x_i$ in each constraint is finite.
{  It has been argued in \cite{MW14} that isostaticity gives
  rise to infinite correlations, and that for this reason  at jamming
  hard spheres behave under stress as a system with long range
  forces. Our hypothesis suggests that this could be the case for a
  large class of systems. 
}

The perceptron is an extreme limit of the hard sphere problem in
infinite dimensions where all the particles except one are pinned in
random positions. It is known that in low dimensional hard sphere systems the
exponents are independent of the fraction of pinned particles
\cite{BPZ13}.  One of the interests of the perceptron is that it is a
much simpler model than hard spheres -even in the infinite dimensional
limit-.  The derivation of the replica effective action is much more
direct and the study with the cavity method would be
straightforward. Though the connection of this model with jamming had
never been underlined before, a lot is known on the model even at the
rigorous level.  Many questions that one could ask about jamming could
be answered more easily in this context than in the hard spheres.  Our
present interests include the computations of spectrum of vibrational
normal modes at small temperature and the low temperature specific
heat in a quantum version of the model.  This is technically feasible:
more surprises are to be waited.

\paragraph*{Acknowledgments --}
We would like to thank G. Biroli, P. Charbonneau, M. Lenz,
M. M\"uller, M. Wyart
for very useful suggestions and  P. Urbani and
F. Zamponi for extremely valuable discussions and for careful reading
of the manuscript. The European Research
Council has provided financial support through ERC grant agreement
no.~247328. SF acknowledges the hospitality of the Physics Departement
of the ``Sapienza'' University of Rome.

\bibliography{BIBLIO}

\section{Supplementary material }

\paragraph{The de Almeida-Thouless line --}
The line of instability of the replica symmetric solution can be
determined from the conditions that a) the RS value of the overlap $q$
is solution of the saddle point equation, b) the l.h.s. of
eq. (\ref{eq:5}), which should be positive in a stable solution,
vanishes on the line, c) the RS solution coincides with a degenerate
RSB solution with $q_0=q_1=q$ and the ``breaking point''
$x_c=\lim_{q_1\to q} x(q_1)$ lies in the interval $[0,1]$. 
The breaking point $x_c$ can be computed from (\ref{eq:16}). Consistency
 requires $x_c<1$; if $x_c$ so identified is $x_c>1$, this is a signal
 that a 1RSB transition occurs before the RS solution becomes
 unstable. In fig. 1 we show that full RSB transition occurs in the
 range $0>\kappa>\kappa_{1rsb}\approx -2.05$ where $x_c<1$, for
 $\kappa<\kappa_{1rsb}$ the found value is  $x_c>1$ and a 1RSB -or
 random first order transition- can be expected.
\begin{figure}
\epsfxsize=250pt 
\epsffile{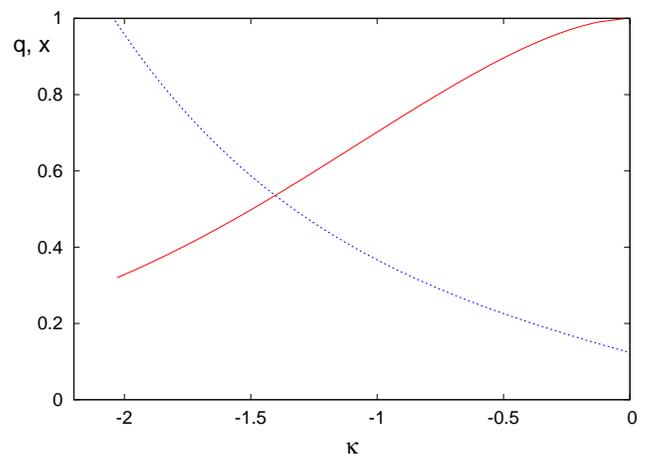}
\caption{Value of the overlap $q$ and of the breaking point $x_c$ on
  the dAT line as a function of $\kappa$. As long as $\kappa<0$, the
  value of $q$ at the RSB transition is less than unity and RSB occurs
  in the SAT phase. In the
  interval $[\kappa_{1RSB},0]$ the transition is continuous. Below that
  value (not shown) a Random First Order transition is to be expected  }
\label{fig3}
\end{figure}

\paragraph{Asymptotic solution close to jamming} {In order to
  characterise the scaling regime close to jamming, where
  $1-q_1\to 0$, let study first the solution to the equations
  (\ref{eq:3},\ref{eq:8}) for large values of $|h|$ and consider the
  equation for $m$ first.  For large negative $h$ one can assume a
  form $m(q,h)=-\frac{h}{ r(q)}$, which inserted into (\ref{eq:8})
  gives \begin{eqnarray} \label{eq:7}
    \dot{r}(q)=-x(q).  \end{eqnarray} The solution that respects the
  condition in $q_1$ is $r(q)=\lambda(q)$.  On the other extreme, for
  $h\to +\infty $ one trivially has $m(q,h)=0$. Let us now study the
  equation for $P(q,h)$. For large positive $h$, one just have a
  diffusion equation, and since for $q\to q_0$ $P(q_0,h)=D_{q_0}h/dh$,
  then $P(q,h)=D_{q}h/dh$, which tends to a Gaussian with unit
  variance for $q\to 1$. For large negative $h$ on the other hand, it
  is simple to show that $P(q,h)$ must have the form $P(q,h)=B(q)
  e^{-A(q)\frac {h^2} 2}$. The factors $A$ and $B$ should
  verify: 
\begin{eqnarray} \label{eq:9} \frac 1 2 \frac{\dot{A}} A =
    \frac{\dot{B}}{B}=-\frac A 2 +\frac x \lambda.  
\end{eqnarray} 
In
  order to solve these equations to the leading order, we should know
  something more about the behaviour of the function $x(q)$.  We are
  interested to the scaling regime when $ 1-q_1\ll 1$. Since
  $x(q)$ is the weight of the overlap lower than $q$ one can expect it
  to go to zero in the jamming limit, we therefore
  write 
\begin{eqnarray} \label{eq:10} x(q)=x_1
  \left(\frac{1-q_1}{\Delta}\right)^{\frac 1 k} 
\end{eqnarray} 
with
  $1-q_1\ll x_1\ll 1$, 
where 
\begin{eqnarray} 
\label{eq:18}
    \lambda(q)\approx \int_q^1 x(q)\approx
    \frac{k}{k-1}x(q)\Delta  
\end{eqnarray} 
and we denoted $\Delta=1-q$. 
In this regime, supposing
  self-consistently $2/k>1$, we can solve the equations (\ref{eq:9})
  supposing that $A\ll x/\lambda \sim \frac{1}{\Delta}$. In this case,
  we find $A(q)\sim \Delta^{-2+2/k}$ and $B(q)\sim
  \Delta^{-1+1/k}$.   We conclude that $P$ and $m$ behave respectively
  as 
$P(q,h)\approx p_2(h)$ is approximately independent of $q$, and
$m(q,h)\approx 0$
 for large positive $h$ while $P(q,h)\approx p_0(h
 \Delta^{-c/k})\Delta^{-c/k}$ with $c=k-1$ and $m(q,h)\approx
 -h/\lambda(q)$ for large negative $h$. These regimes should be
 matched by functions $\Delta^{-a/k}p_1(h\Delta^{-b/k})$ (with $a< b$)
 and $\frac{\Delta^{b/k}}{\lambda(q)} m_1(h\Delta^{-b/k})$.
 Notice that the in the jamming limit $q\to 1$, 
\begin{eqnarray}
 P(q,h)\to A \delta(h)+p_2(h) 
\end{eqnarray}
with $A=\int du \; p_0(u)$. Apart for the
$\delta$-function 
term, $p_2(h)$ is the physical distribution of the gaps at jamming and
should scale as 
$p_2(h)\sim h^{-\gamma}$ at small argument.  We notice at this point
that for $q
\to 1$, eq. (\ref{eq:5})
 reduces to
 \begin{eqnarray}
   \label{eq:14}
1 ={\alpha}\int_{-\infty}^0 dh\; p_0(h)
 \end{eqnarray}
which gives $A=1/\alpha$ i.e. the condition that the system is isostatic. 
 Generalising the
thermodynamic calculation to compute 
the ground state energy slightly above jamming, 
for $\kappa>\kappa_c(\alpha)$, one sees that, apart from a 
multiplicative constant, the negative part of the distribution,
$p_0(-f)$ can be identified with the force distribution
\cite{CKPUZ14}.
We can therefore assume a power scaling $p_0(u)\sim |u|^\theta$ at small argument,
  to be matched with the behaviour of $p_1(z)$ at large negative $z$,
  analogously, the behaviour $p_2(h)\sim |h|^{-\gamma}$ at small
  positive argument should be matched with the behaviour of $p_1$ for
  large positive $z$. Consistency requires that
  $\theta=\frac{c-a}{b-c}$, $c=k-1$, $\gamma=a/b$ and $b/k=1/2$.  
We find at this point the remarkable result that the equations
  and boundary conditions that determine  $p_1$ and $m_1$ coincide with the
  corresponding ones in the case of jamming of hard spheres.
Writing them explicitly we have:
\begin{eqnarray}
  \label{eq:11}
  && \frac a k p_1+\frac 1 2 z p_1'=\frac 1 2 \left (
    p_1''+2 \frac c k(m_1p_1)'\right)\\
&& (\frac c k -\frac 1 2 )m_1+\frac 1 2 z m_1' = -\frac 1 2
\left( m_1'' +2 \frac c k m_1m_1'\right)
  \label{eq:11bis}
\end{eqnarray}
with boundary conditions 
\begin{eqnarray}
  \label{eq:12}
  &&p_1(z)=\left\{
    \begin{array}{ll}
     z^\theta & z\to +\infty\\
      z^{-\gamma} & z\to -\infty
    \end{array}
\right. \\
\label{eq:12bis}
  &&m_1(z)=\left\{
    \begin{array}{ll}
     0 & z\to +\infty\\
      -z & z\to -\infty
    \end{array}
\right. 
\end{eqnarray}
In addition the solution should verifies
\begin{eqnarray}
  \label{eq:22}
  &&\int du \; p_1(u) \left[
(k-1)m_1'(u)^2(1+m'(u))-k m_1''(u)^2 
\right]\nonumber\\&&=0 
\end{eqnarray}
that follows from eq. (\ref{eq:5}) and its derivative w.r.t. $q$ in
the scaling domain.  
 We can just then quote from \cite{CKPUZ14} the values of $\gamma=0.41269$,
and $\theta=0.42311$. We should note at this point that while the
exponents and the scaling functions $p_1$ and $m_1$ appear to be
universal, the functions $p_0$ and $p_2$, apart for their small
argument part, are system specific and non universal. 
\paragraph{Random Linear Programming} 
The replicated entropy of the RLP model is very similar to the one
of the perceptron, indeed formula (\ref{srep}) is substituted by:
\begin{eqnarray}
  n S[Q]=\frac 1 2 \tr \;\log \; Q + \\ \alpha \log \left.\left( e^{ \frac 1 2 \sum_{ab}[\sigma+ Q_{ab}] \frac{\partial ^2}{\partial h_a\partial h_b} }\prod_a \theta(h_a-\kappa)\right)\right|_{h_a=0}\nonumber
\label{srep1}
\end{eqnarray}
with the important difference that here since there is no spherical
constraint the diagonal value of $Q_{ab}$, $Q_{aa}=\tilde{q}=\frac 1 N
\sum_{i=1}^N {x_i^a}^2$ is a variational parameter to be determined by
the saddle point equations. Since the problem is convex replica
symmetry always holds, and 
  \begin{eqnarray}
  \label{eq:16} && S=\frac 1 2
\left[\log(\tilde{q}-q)+\frac{q}{\tilde{q}-q}\right]+\\ &&\alpha \int
D_{\sigma+q} y \; \log H\left( \frac{\kappa-y}{\sqrt{\tilde{q}-q}}
\right)
\nonumber
\end{eqnarray}
in the jamming limit in which $\Delta=\tilde{q}- q\to 0$, $S$ should
not diverge faster than $\log \Delta$. Expanding
$S$ in $\Delta$ we have
 \begin{eqnarray}
 S\approx \frac {1}{2 \Delta}
\left[
q-\alpha \int
D_{\sigma+q} y \;(y-\kappa)^2\theta(\kappa-y) \right]
\end{eqnarray}
both this term and its $q$-derivative should vanish; we get
\begin{eqnarray}
  \label{eq:21}
&& 0=  q-\alpha \int 
D_{\sigma+q} y \;(y-\kappa)^2\theta(\kappa-y) \\
&& 0=  1-\alpha \int 
D_{\sigma+q} y \; \theta(\kappa-y),
\label{eq:21.2}
\end{eqnarray}
which have a finite $q$ and $\kappa$ solution if $\alpha>2$. 
Eq. (\ref{eq:21.2}) is the RS  version of the condition of marginal
stability (\ref{eq:5}) and
implies isostaticity. From an explicit computation one sees  that the
field distribution is given by 
\begin{eqnarray}
  \label{eq:15a}
  g(r)=\delta(r)+\frac{\alpha}{\sqrt{2 \pi (\sigma+q)}} e^{-(r+\kappa)^2/2(\sigma+q)}\theta(r).
\end{eqnarray}
 We see that the solution
is marginally stable and isostatic, but $g(r)$ is regular in the origin. The
singular behaviour of $g(r)$ appears critically associated to replica
symmetry breaking. This requires an interpretation which is at present
lacking. 

\paragraph{Stability Bound} 
It is possible to obtain the relation between the force and the gap
exponents $\gamma\ge 1/(2+\theta)$ in the perceptron, generalising the
argument of stability used by Wyart in hard spheres \cite{W12}.  To
this aim let us note that for fixed $\alpha$ the maximisation of
$\kappa$ on the sphere with fixed ${\bf x}^2$ is equivalent to the
dual optimisation problem of finding an the maximum (resp. minimum)
value in free space $R^N$ of ${\bf x}^2$ at fixed value of $\kappa$
negative (resp. positive). Let us consider the most interesting case
$\kappa<0$.  The core of the argument is that in isostatic
configurations, in absence of small enough gaps one could increase the
value of ${\bf x}^2$ just by following the 'floppy mode' that ensue
from the opening of a contact; stability then requires that small
forces be in correspondence with small enough gaps.  

The maximisation
of ${\bf x}^2$ under the perceptron constraints can be performed
introducing Karush-Kuhn-Tucker multipliers $F^\mu$ (forces) and the
objective function:
\begin{eqnarray}
  \label{eq:23}
  L({\bf x})=x^2+\sum_{\mu=1}^{\alpha N} F^\mu[\frac{1}{\sqrt{N}}{\bf x}\cdot
  \xi^\mu-\kappa]. 
\end{eqnarray}
The first order maximisation conditions read: 
\begin{eqnarray}
  \label{eq:24}
  \frac {\partial L}{\partial x_i}=2 x_i+\frac{1}{\sqrt{N}} \sum_{\mu}F^\mu\xi_i^\mu=0. 
\end{eqnarray}
In an isostatic maximum, there are $N$ positive forces $F^\mu$ in
correspondence with the binding
constraints while all the others are zero. Let us suppose without loss of
generality to renumber the constraints so that the
$0<F^1<F^2<...<F^{N}$ while $F^\mu=0$ for $\mu>N$, and study the
effect of unbinding the first contact by an extent $s$. We change then
${\bf x}$ into ${\bf x}+\delta {\bf x}$ in such a way
$\frac{1}{\sqrt{N}} ({\bf x}+\delta {\bf x})\cdot
\xi^1=\kappa+s$, while keeping the remaining contacts binding:
\begin{eqnarray}
  \label{eq:25}
\frac{\delta {\bf x}\cdot
\xi^\mu}{\sqrt{N}} = s \delta_{\mu,1}\;\;\; {\rm for} \;\;\;\mu=1,...,N  
\end{eqnarray}
Eq. (\ref{eq:25}) has a unique solution and implies that each of the $\delta x_i$  for
$i=1,...,N$ is of order $s$. Using
(\ref{eq:24},\ref{eq:25}), the variation of ${\bf x}^2$, $\Delta {\bf x}^2=2 {\bf
  x}\cdot \delta {\bf x}+(\delta {\bf x})^2$  can be then written as 
\begin{eqnarray}
  \label{eq:26}
  \Delta {\bf x}^2=-F^1s +A N s^2
\end{eqnarray}
where $A$ is a constant of order 1.  Notice that $\Delta {\bf x}^2<0$
for small $s$, while it changes sign for $s=s^*\equiv F^1/(AN)$. The
maximum is stable if before reaching that point, a new contacts forms,
that prevents further maximisation of ${\bf x}^2$.  At this point the
argument proceeds verbatim as in \cite{W12}, we reproduce it here just
for completeness. The less restrictive assumption one can make on the
force distribution is the presence of a power singularity $P(F)\sim
F^\theta$, which implies $F^1=F_{min}\sim
N^{-\frac{1}{1+\theta}}$. Opening the weakest contact would therefore
imply a growth in ${\bf x}^2$ if $s>s^*\sim
N^{-1-\frac{1}{1+\theta}}$. However, a new contact is formed and blocks
the floppy mode at a value of $s$ of the order of $r_{min}$ the
minimum gap in the system. Stability requires therefore
$r_{\min}{<\atop\sim} s^*$.  Consequently, the distribution $g(r)$
should be power law in the origin $g(r)\sim r^{-\gamma}$ in such a way
that $r_{\min}\sim N^{-\frac{1}{1-\gamma}}{<\atop\sim}
N^{-1-\frac{1}{1+\theta}}$. The inequality $\gamma\ge 1/(2+\theta)$
follows readily.  Two facts should be noted:
\begin{enumerate}
\item If
$\kappa>0$ ($\alpha<2$) maximisation of $\kappa$ at fixed ${\bf x}^2$
is dual (equivalent) to {\it minimisation} of ${\bf x}^2$ at fixed
$\kappa$. In this case, the forces are negative, and both terms in
(\ref{eq:26}) lead to an increase of ${\bf x}^2$. 
\item {  For the RLP $\kappa$ maximisation is by no means equivalent to ${\bf
    x}^2$ maximisation. }
\end{enumerate}
In both cases, small forces do not need to be compensated by the
existence of small gaps and there is no necessarily a relation between
the two distributions.
\end{document}